\documentclass[preprint,showpacs,preprintnumbers,amsmath,amssymb]{revtex4}
\usepackage{graphicx}
\usepackage{dcolumn}
\usepackage{bm}

\newcommand{\be}{\begin{equation}}
\newcommand{\en}{\end{equation}}
\newcommand{\bea}{\begin{eqnarray}}
\newcommand{\ena}{\end{eqnarray}}

\begin{document}


\title{ Warm-Intermediate  inflationary universe model in braneworld cosmologies }
\author{Ram\'on Herrera}
\email{ramon.herrera@ucv.cl} \affiliation{ Instituto de
F\'{\i}sica, Pontificia Universidad Cat\'{o}lica de
Valpara\'{\i}so, Casilla 4059, Valpara\'{\i}so, Chile.}

\author{Eugenio San Martin}
 \email{eugenio.sanmartin.r@mail.pucv.cl }

\affiliation{ Instituto de F\'{\i}sica, Pontificia Universidad
Cat\'{o}lica de Valpara\'{\i}so, Casilla 4059, Valpara\'{\i}so,
Chile.}

\date{\today}

\begin{abstract}
 Warm-intermediate inflationary universe models in the context
 of  braneworld cosmologies,  are studied.   This study is done in the weak and
strong dissipative  regimes. We find that, the scalar potentials
and dissipation coefficients in terms of the scalar field, evolves
as type-power-law and powers of logarithms, respectively. General
conditions required for these models to be realizable are derived
and  discussed. We also study the scalar and tensor perturbations
for each  regime. We use recent astronomical observations  to
constraint  the parameters appearing in the braneworld models.
\end{abstract}

\maketitle

\section{Introduction}

It is well  know that warm inflation, as opposed to the
conventional cool inflation, presents the attractive feature that
it avoids the reheating period \cite{warm,agre}. The warm
inflation scenario differs from the cold inflation scenario in
that there is no separate reheating phase in the former, and
rather radiation production occurs concurrently with inflationary
expansion. In this way, warm inflation provides an alternative to
the traditional reheating
 mechanism by smoothly connecting an early inflationary period with a
radiation dominated phase. In these kind of models dissipative
effects are important during the inflationary period, so that
radiation production occurs concurrently together with the
inflationary expansion. If the radiation field is in a highly
excited state during inflation, and this has a strong damping
effect on the inflaton dynamics, then, it is found a strong
regimen of  warm inflation. Also, the dissipating effect arises
from a friction term which describes the processes of the scalar
field dissipating into a thermal bath via its interaction with
other fields. Warm inflation shows how thermal fluctuations during
inflation may play a dominant role in producing the initial
fluctuations  necessary for Large-Scale Structure (LSS) formation.
In this way, density fluctuations arise from thermal rather than
quantum fluctuations \cite{62526,1126}.  Among the most attractive
features of these models, warm inflation end  at the epoch when
the universe stops inflating and "smoothly" enters in a radiation
dominated Big-Bang phase\cite{warm}. The matter components of the
universe are created by the decay of either the remaining
inflationary field or the dominant radiation field
\cite{taylorberera}.

On the other hand, a possible evolution during inflation is the
particular scenario of intermediate inflation, in which the scale
factor, $a(t)$, evolves as $a=\exp(A t^f)$, where $A$ and $f$ are
two constants, where $0<f<1$; the expansion of this universe is
slower than standard de Sitter inflation ($a=\exp(H t)$), but
faster than power law inflation ($a= t^p; p>1$)\cite{atp}, this is
the reason why it is called "intermediate". This model was
introduced as an exact solution for a particular scalar field
potential of the type $V(\phi)\propto
\phi^{-4(f^{-1}-1)}$\cite{Barrow1}. In the slow-roll
approximation, and with this sort of potential, it is possible to
have a spectrum of density perturbations which presents a
scale-invariant spectral index, i.e. $n_s=1$, the so-called
Harrizon-Zel'dovich spectrum provided that $f$ takes the value of
$2/3$\cite{Barrow2}. Even though this kind of spectrum is
disfavored by the current WMAP data\cite{astro,astro2}, the
inclusion of tensor perturbations, which could be present at some
point by inflation and parametrized by the tensor-to-scalar ratio
$r$, the conclusion that $n_s \geq 1$ is allowed providing  that
the value of $r$ is significantly nonzero\cite{ratio r}. In fact,
in Ref. \cite{Barrow3} was shown that the combination $n_s=1$ and
$r>0$ is given by a version of the intermediate inflation  in
which the scale factor varies as $a(t)\propto e^{t^{2/3}}$ within
the slow-roll approximation. Eventually, warm inflationary
universe models in the context of intermediate inflation in
General Relativity was studied in Ref.\cite{yo4} and inflation
intermediate on the brane was considered in Ref.\cite{yo6}, where
the value of the tensor-scalar ratio $r$ is significantly nonzero.

In view of these observations, it is important to further our
understanding of the inflationary models from a theoretical
perspective. Therefore, there is considerable interest in
inflationary scenarios  motivated by superstring and M-theory (
see Refs.\cite{6,7}). Specifically, the braneworld scenario, where
our observable, four–dimensional universe is regarded as a domain
wall embedded in a higher–dimensional bulk space \cite{8}.  This
theory suggests that in order to have a ghost-free action high
order curvature invariant corrections to the Einstein-Hilbert
action must be proportional to the Gauss-Bonnet (GB)
term{\cite{BD}}. The GB terms arise naturally as the leading order
of the $\alpha$ expansion to the low-energy string effective
action, where $\alpha$ is the inverse string tension{\cite{KM}}.
This kind of theory has been applied to possible resolution of the
initial singularity problem{\cite{ART}}, to the study of
Black-Hole solutions{\cite{ Varios1}}, accelerated cosmological
solutions{\cite{ Varios2}}. Recently, accelerated expansion in  an
intermediate inflationary universe models using the GB brane was
studied in Ref.\cite{yo5} and in a closed or open inflationary
universe models using the GB brane was considered in
Ref.\cite{yo7}. In particular, in Ref.{\cite{Sanyal}}, it has been
found that for a dark energy model the GB interaction in four
dimensions with a dynamical dilatonic scalar field coupling leads
to a solution of the form $a\propto \exp{A t^{f}}$, where the
universe starts evolving with a decelerated exponential expansion.

In this way, the idea that inflation, or specifically,
intermediate inflation, comes from an effective theory at low
dimension of a more fundamental string theory is in itself very
appealing. Thus, our aim in this paper is to study an evolving
intermediate scale factor in the warm inflationary universe
scenario in the context  of  braneworld cosmologies. We will do
this for two regimes; the weak and the strong dissipative regimes.

The outline of the paper is a follows. The next section presents a
short review of  the modified  Friedmann equation and the
warm-intermediate inflationary phase in braneworld cosmologies. In
the Sections \ref{section2} and \ref{section3} we discuss the weak
and strong dissipative regimes, respectively. Here, we give
explicit expressions for the dissipative coefficient, the scalar
power spectrum and the tensor-scalar ratio. Finally, our
conclusions are presented in Section\ref{conclu}.  We chose units
so that $c=\hbar=1$.

\section{The  Warm-Intermediate Inflationary phase in Gauss Bonnet.\label{secti} }

We start with the five-dimensional bulk action for the GB
braneworld:

\begin{eqnarray}
S &=&\frac{1}{2\kappa
_{5}^{2}}\int_{bulk}d^{5}x\sqrt{-g_{5}}\left\{ R-2\Lambda
_{5}+\alpha \left( R^{\mu \nu \lambda \rho }R_{^{\mu \nu \lambda
\rho }}-4R^{\mu \nu }R_{\nu \mu }+R^{2}\right) \right\} \nonumber \\
&&+\int_{brane}d^{4}x\sqrt{-g_{4}}\left(
\mathcal{L}_{matter}-\sigma \right),\label{action1}
\end{eqnarray}
where $\Lambda _{5}=-3\mu ^{2}\left( 2-4\alpha \mu ^{2}\right) $
is the cosmological constant in five dimensions, with the
$AdS_{5}$ energy scale $\mu$, $\alpha$ is the GB coupling
constant, $\kappa _{5}=8\pi/m_{5}$ is the five dimensional
gravitational coupling constant and $\sigma$ is the brane tension.
 $\mathcal{L}_{matter}$ is the matter lagrangian for the
inflaton field on the brane. We will consider the case that a
perfect fluid matter source with density $\rho$ is confined to the
brane.

A Friedmann-Robertson-Walker (FRW) brane in an AdS$_5$ bulk is a
solution to the field and junction equations (see Refs.\cite{Ch,
Davis:2002gn, Gravanis:2002wy}). The modified Friedmann on the
brane can be written as
 \begin{equation}
H^2 = {1\over
4\alpha}\left[(1-4\alpha\mu^2)\cosh\left({2\chi\over3}
\right)-1\right]\,,\label{q22}
 \end{equation}
\begin{equation}
\label{var1} \kappa_5^2(\rho+\sigma) =
\left[{{2(1-4\alpha\mu^2)^3} \over {\alpha} }\right]^{1/2}
\sinh\chi\,,
\end{equation}
where $\chi$ represents a dimensionless measure of the energy
density $\rho$.
In this work we will assume that the matter fields are restricted
to a lower dimensional hypersurface (brane) and that gravity
exists throughout the space-time (brane and bulk) as a dynamical
theory of geometry. Also, for 4D homogeneous and isotropic
Friedmann cosmology, an extended version of Birkhoff's theorem
tells us that if the bulk space-time is AdS, it implies that  the
effect of the Weyl tensor (known as dark radiation) does not
appear in the modified Friedmann equation. On the other hand, the
brane Friedmann equation for the general,  where the bulk
space-time may be interpreted as a charged black hole was studied
in Refs.\cite{Od,Od2,Od3}.

The modified Friedmann equation~(\ref{q22}), together with
Eq.~(\ref{var1}), shows that there is a  characteristic GB energy
scale\cite{pp}, where
$m_{GB}= \left[{{2(1-4\alpha\mu^2)^3} \over {\alpha} \kappa_5^4
}\right]^{1/8}\,$,
such that the GB high energy regime ($\chi\gg1$) occurs if
$\rho+\sigma \gg m_{GB}^4$.  Expanding Eq.~(\ref{q22}) in $\chi$
and using Eq.(\ref{var1}), we find in the full theory three
regimes for the dynamical history of the brane universe \cite{Ch,
Davis:2002gn, Gravanis:2002wy}:

 \be
\rho\gg m_{GB}^4~ \Rightarrow ~ H^2\approx \left[ {\kappa_5^2
\over 16\alpha}\, \rho
\right]^{2/3}\,\;\;\;\;\;\;\;(GB),\label{gbl}
 \en
 \be
 m_{GB} \gg
\rho\gg\sigma  \Rightarrow ~ H^2\approx {\kappa_4^2 \over
6\sigma}\, \rho^{2}\,\;\;\;\;\;\;\;\;\;\;(RS),\label{rsl}
 \en
 \be
\rho\ll\sigma~ \Rightarrow ~ H^2\approx {\kappa_4^2 \over 3}\,
\rho\,\;\;\;\;\;\;\;\;\;\;\;\;\;\;\;\;\;\;\;\;\;\;\;(GR).
\label{ehl}
 \en
Clearly Eqs. (\ref{gbl}), (\ref{rsl}) and (\ref{ehl}) are much
simpler than the full Eq (\ref{q22}) and in a practical case one
of the three energy regimes will be assumed. Therefore,  can be
useful to describe the universe in a region of time and energy in
which \cite{Calcagni:2004bh,Tsujikawa:2004dm}
\begin{equation}
H^{2}=\beta_{q}^2\rho ^{q}, \label{dda}
\end{equation}
where $H=\dot{a}/a$ is the Hubble parameter  and $q$ is a patch
parameter that describes a particular cosmological model under
consideration. The choice $q=1$ corresponds to the standard
General Relativity (GR) with $\beta^2_1=8\pi/3m_{p}^2=\kappa^2/3$,
where $m_{p}$ is the four dimensional Planck mass. If we take
$q=2$, we obtain the high energy limit of the brane world
cosmology, Randall-Sundrum (RS), in which $\beta^2_2=4\pi/3\sigma
m_p^2 =\kappa^2/6\sigma$. Finally, for $q=2/3$, we have the GB
brane world cosmology, with $\beta^2_{2/3}=G_{5}/16\zeta$, where
$G_5$ is the $5D$ gravitational coupling constant and
$\zeta=1/8g_s$ is the GB coupling ($g_s$ is the string energy
scale). The parameter $q$, which describes the effective degrees
of freedom from gravity, can take a value in a non-standard set
because of the introduction of non-perturbative stringy effects.
Here, we mentioned some possibilities, for instance, in
Ref.\cite{Kim:2004hu} it was found that an appropriate region to a
patch parameter $q$ is given by $1/2 = q < \infty$. On the other
hand, from Cardassian cosmology it is possible to obtain a
Friedmann equation similar  (\ref{dda}) as a consequence of
embedding our observable universe as a 3+1 dimensional brane in
extra dimensions. In particular,  a modified FRW equation  was
obtained in our observable brane with $H^2 \propto \rho^n$ for any
$n$ in Ref.\cite{Chung:1999zs}.


On the other hand, we neglect any contribution from both the Weyl
tensor and the brane-bulk exchange, assuming there is some
confinement mechanism for a perfect fluid. Thus, the energy
conservation equation on the brane follows directly from the
Gauss-Codazzi equations. For a perfect fluid matter source it is
reduced to the familiar form, $ \dot{\rho}+3H\left( \rho +P\right)
=0,$  where $P$ represent the  pressure density. The dot denotes
derivative with respect to the cosmological time $t$.

In the following we will consider  a total energy density
$\rho=\rho_\phi+\rho_\gamma$, where $\phi$ corresponds to a
self-interacting scalar field with energy density, $\rho_\phi$,
given by $\rho_\phi=\dot{\phi}^2/2+V(\phi)$,
  $V(\phi)=V$ is
the scalar   potential and  $\rho_\gamma$ represents  the
radiation energy density.

From Eq.(\ref{dda}), we assume that the gravitational dynamics
give rise to a Friedmann equation of the form
\begin{equation}
H^2=\beta_q^2\,[\rho_{\phi}+\rho_\gamma]^q. \label{HC}
\end{equation}

 The dynamics of the
cosmological model, for $\rho_\phi$ and $\rho_\gamma$ in the warm
inflationary scenario is described by the equations
 \be\dot{\rho_\phi}+3\,H\,(\rho_\phi+P_\phi)=-\Gamma\;\;\dot{\phi}^2, \label{key_01}
 \en
and \be \dot{\rho}_\gamma+4H\rho_\gamma=\Gamma\dot{\phi}^2
.\label{3}\en Here $\Gamma$ is the dissipation coefficient and it
is responsible of the decay of the scalar field into radiation
during the inflationary era. $\Gamma$ can be assumed to be a
constant or a function of the scalar field $\phi$, or the
temperature $T$, or both \cite{warm}.  On the other hand, $\Gamma$
must satisfy  $\Gamma>0$ by the Second Law of Thermodynamics.

During the inflationary epoch the energy density associated to the
scalar field dominates over the energy density associated to the
radiation field\cite{warm,62526}, i.e., $\rho_\phi>\rho_\gamma$
and  the Friedmann equation (\ref{HC})  reduces  to
\begin{eqnarray}
H^2\approx\beta_q^2\;\rho_\phi^q,\label{inf2}
\end{eqnarray}
and  from Eqs. (\ref{key_01}) and (\ref{inf2}), we can write
\begin{equation}
 \dot{\phi}^2= -\frac{2\,H ^{\frac{2}{q}-2}\dot{H}}{3\;{q}\;\beta_q\;^\frac{2}{q}\,(1+R)},\label{inf3}
\end{equation}
where $R$ is the rate defined as
\begin{equation}
 R=\frac{\Gamma}{3H }.\label{rG}
\end{equation}
For the  weak (strong) dissipation  regime, we have $R< 1$ ($R>
1$) (see Refs.\cite{warm,62526}).

We also consider that  during  warm inflation the radiation
production is quasi-stable\cite{warm,62526}, i.e.
$\dot{\rho}_\gamma\ll 4 H\rho_\gamma$,\,  $
\dot{\rho}_\gamma\ll\Gamma\dot{\phi}^2$ and from Eq.(\ref{3}) we
obtained that
 \begin{equation}
\rho_\gamma=\frac{\Gamma\dot{\phi}^2}{4H}=-\frac{\Gamma\,H^{\frac{2}{q}-3} \dot{H}}{6q\,\beta_q\,^{\frac{2}{q}}\,(1+R)},\label{rh}
\end{equation}
which  could be written as $\rho_\gamma= C_\gamma\, T^4$, where
$C_\gamma=\pi^2\,g_*/30$ and $g_*$ is the number of relativistic
degrees of freedom. Here $T$ is the temperature of the thermal
bath.

From Eqs.(\ref{inf3}) and (\ref{rh}) we get that
\begin{equation}
T= \left[-\frac{\Gamma\,H\,^{\frac{2}{q}-3}\dot{H}}{6\,\,\,C_\gamma
\,q\,\beta_q^{\frac{2}{q}}(1+R)}\right]^{1/4}.\label{rh-1}
\end{equation}

On the other hand, in warm inflation the interactions are
important during the inflationary scenario. If the fields
interacting with the inflaton are at high temperature, then it is
complex to control the thermal loop corrections to the effective
potential that is required  to preserve the appropriate  flat
potential required for inflation. Nevertheless, if the fields
interacting with the inflaton are at low temperature, then
supersymmetry can be applied  to cancel the quantum radiative
corrections, and maintain an appropriate  potential \cite{26}.

From first principles in quantum field theory the dissipation
coefficient $\Gamma$ is computed for models in cases of
low-temperature regimes\cite{26} (see also Ref.\cite{28}). Here,
was developed the dissipation coefficients in supersymmetric
models which have an inflaton together with multiplets of heavy
and light fields. In this approach, it was used   an interacting
supersymmetric theory, which has three superfields $\Phi$, $X$ and
$Y$ with a superpotential, $W=g\Phi X^2+hXY^2$, where $\phi$,
$\chi$ and $y$ refer to their bosonic component.  The interaction
structure for this superpotential is habitual in many particle
physics SUSY models during inflation. Also, this superpotential
can simply be modified to develop a hybrid inflationary model. The
inflaton field couples to heavy bosonic field $\chi$ and fermions
$\psi_\chi$, obtain their masses through couplings to $\phi$,
where $m_{\psi_\chi}=m_\chi=g\phi$. In the low -temperature
regime, i.e. $m_\chi,m_{\psi_\chi}>T>H$, the dissipation
coefficient, when $X$ and $Y$ are singlets, becomes \cite{26}
\begin{equation}
\Gamma\simeq0.64\,g^2\,h^4\left(\frac{g\,\phi}{m_\chi}\right)^4\,
\frac{T^3}{m_\chi^2}.\label{G0}
\end{equation}
This latter equation can be rewritten as
\begin{equation}
\Gamma\simeq C_\phi\,\frac{T^3}{\phi^2},\label{G}
\end{equation}
where $C_\phi=0.64\,h^4\,\cal{N}$. Here ${\cal{N}}={\cal{N}}_\chi
{\cal{N}}_{decay}^2$, where $\cal{N}_\chi$ is the multiplicity of
the $X$ superfield and ${\cal{N}}_{decay}$ is the number of decay
channels available in $X$'s decay\cite{26,27}.

From Eq.(\ref{rh-1}) the above equation becomes
\begin{equation}
\Gamma^{1/4}\,(1+R)^{3/4}\simeq\left[-\frac{\,H^{\frac{2}{q}-3}\,\dot{H}}{6\,
C_\gamma\,q\,\beta_q^{\frac{2}{q}}}\right]^{3/4}\,\frac{C_\phi}{\phi^2},\label{G1}
\end{equation}
which determines the dissipation coefficient in the weak (or
strong) dissipative regime  in terms of scalar field $\phi$ and
the parameters of the model.

 We should note that in general the scalar potential  from
Eqs.(\ref{HC}) and (\ref{rh}) becomes
\begin{equation}
V(\phi)=\left[\frac{H}{\beta_q}\right]^{\frac{2}{q}}\left[1 \, +\frac{\dot{H}}{3\,q\,H^{2}(1+R)}\,\left(1+\frac{3}{2}\,R\right)\right],\label{pot}
\end{equation}
which could be expressed explicitly in terms of the scalar field,
$\phi$, by using Eqs.(\ref{inf3}) and (\ref{G1}), in the
 weak (or strong) dissipative regime.

On the other hand, solutions can  be found for warm-intermediate
inflationary universe models where  the scale factor, $a(t)$,
expands as follows
\begin{equation}
a(t)=\exp(\,A\,t^{f}).\label{at}
\end{equation}

In the following, we develop models for a variable dissipation
coefficient $\Gamma$, and  we will restrict ourselves to the weak
(or strong ) dissipation regime.

\section{ The  weak dissipative regime.\label{section2}}

Assuming that, once the system evolves according to the weak
dissipative regime, i.e. $\Gamma<3H$, it remains in such limit for
the rest of the evolution. From Eqs.(\ref{inf3}) and (\ref{at}),
we obtained a relation between the scalar field and cosmological
times given by
\begin{equation}
\phi(t)=\phi_0+\frac{C}{2\theta}\,t^{\theta},\label{wr1}
\end{equation}
where
$
C=\left[\frac{8\,(Af)^{\frac{2}{q}-1}\,(1-f)}{3q\,\beta_q^{\frac{2}{q}}}\right]^{1/2},$ and
$\theta =\frac{1}{2}\left[(f-1)(\frac{2}{q}-1)+1\right].
$

Here $\phi(t=0)=\phi_0$.  The Hubble parameter as a function of
the inflaton field, $\phi$, results in
\begin{equation}
H(\phi)=A\,f\,\left[\frac{2\,\theta}{C}\right]^{(f-1)/\theta}\,(\phi-\phi_0)^{(f-1)/\theta}.\label{HH}
\end{equation}
Without loss of generality $\phi_0$ can be taken to be zero.

From Eq.(\ref{G1}) we obtain for the dissipation coefficient as
function of scalar field
\begin{equation}
\Gamma(\phi)=D\,\phi^{-\alpha_1},
\end{equation}
where
$
D=(Af)^{6(\frac{1}{q}-1)}C_\phi^{4}\left\{\left[\frac{(1-f)}
{6\,C_\gamma \, q\,\beta_q^{\frac{2}{q}}}\right]\left[\left(\frac{2\,\theta}{C}\right)^{\frac{2(\frac{1}{q}-1)(f-1)-1}{\theta}}\right]\right\}^{3},$ and
$\alpha_1=2+\frac{3}{\theta}\,(f+1).
$

Using the slow-roll approximation, $\dot{\phi}^2/2<V(\phi)$, and
$V(\phi)>\rho_\gamma$, the scalar potential given by
Eq.(\ref{pot}) reduces to
\begin{equation}
V(\phi)\simeq \left(\frac{H}{\beta_q}\right)^\frac{2}{q}=V_0\,\phi^{-\alpha_2},\label{pot11}
\end{equation}
where the constants $V_0$ and $\alpha_2$ are;
$
V_0=\left[
\frac{A\,f\,(\frac{2\,\theta}{C})^{\frac{f-1}{\theta}}}{\beta_q}\right]^{2/q}$ and $\alpha_2=\frac{2(1-f)}{q\,\theta}
$, respectively.
Note that this kind of potential does not present a minimum.  Note
also that the scalar field $\phi$, the Hubble parameter $H$, and
the scalar potential $V(\phi)$ become independent of the parameters
$C_\phi$ and $C_\gamma$.

Introducing the Hubble slow-roll parameters
$(\epsilon_{1},\eta_{\eta})$ and potential slow-roll parameters
$(\epsilon_{1}^{q},\eta_{n}^{q})$, see Ref.\cite{Kim:2004gs}, we
write

\begin{equation}
\epsilon_{1}=-\frac{\dot{H}}{H^{2}}=\epsilon_{1}^{q}=\frac
{qV^{\prime^{2}}}{6\beta_{q}^{2}V^{q+1}}=\frac{(1-f)}{A\,f}\left(\frac{C}{2\,\theta}\right)^{f/\theta}\;\frac{1}{\phi^{f/\theta}},\label{ep}
\end{equation}
 \begin{equation}
\eta_{n}    =-\frac{1}{H^{n}\dot{\phi}}\frac{d^{n+1}\phi}{dt^{n+1}}%
=\eta_{n}^{q},
\end{equation}
and in particular
$$
\eta_{1}^{q}    =\frac{1}{3\beta_{q}^{2}}\left[  \frac{V^{\prime\prime}%
}{V^{q}}-\frac{qV^{\prime^{2}}}{2V^{q+1}}\right]  =\frac{(2-f)}{A\,f}\;\left(\frac{C}{2\,\theta}\right)^{f/\theta}\frac{1}{\phi^{f/\theta}}\,,
$$
and
$$
 \eta_{2}^{q}  =\frac{-1}{(3\beta_q^2)^2}\left[\frac{V'V''}{V^{2q}}+\frac{(V'')^2}{V^{2q}}-\frac{5qV''(V')^2}{V^{2q+1}}
 +\frac{q(q+2)(V')^4}{2\,V^{2(q+1)}}\right].
$$

So, the condition for inflation to occur  $\ddot{a}>0$ (or
equivalently $\epsilon_1^q<$1)   is only satisfied when
$\phi^{f/\theta}>\frac{(1-f)}{A\,f}\left(\frac{C}{2\,\theta}\right)^{f/\theta}$.


The number of e-folds between two different values of cosmological
times $t_1$ and $t_2$ (or equivalently between two values $\phi_1$
and $\phi_2$ of the scalar field)   is given by
\begin{equation}
N=\int_{t_1}^{t_{2}}\,H\,dt=A\,(t_{2}^f-t_1^f)=A\,\left(\frac{2\,\theta}{C}\right)^{f/\theta}
(\phi_{2}^{f/\theta}-\phi_1^{f/\theta}).\label{N1}
\end{equation}
Here we have used Eq.(\ref{wr1}).

If we assume that inflation begins at the earliest possible stage
(see Ref.\cite{Barrow3}), that it, at $\epsilon_1^q=1$ (or
equivalently $\ddot{a}=0$ ), the scalar field becomes
\begin{equation}
\phi_{1}=\left(\frac{\,1-f}{A\,f}\right)^{\theta / f}\left(\frac{C}{2\,\theta}\right)\;. \label{al}
\end{equation}


On the other hand, as argued in Refs.\cite{warm,Liddle}, the
amplitude of scalar perturbations generated during inflation for a
flat space is approximately
${\cal{P}_{\cal{R}}}^{1/2}=\frac{H}{\dot{\phi}}\,\delta\phi$. In
particular in the warm inflation regime, a thermalize radiation
component is present, therefore, inflation fluctuations are
dominantly thermal rather than quantum.  In the weak dissipation
limit, we have $\delta\phi^2\simeq H\,T$ \cite{62526,B1}. From
Eqs.(\ref{inf3}) and (\ref{rh-1}),  ${\cal{P}_{\cal{R}}}$ becomes
\begin{equation}
{\cal{P}_{\cal{R}}}\simeq\left[\frac{3^{3}\,q^{3}\,\beta_q^{\frac{6}{q}}\,\Gamma}{2^5\,C_\gamma}\right]^{1/4}\;
\left[\frac{H^{\frac{17}{3}-\frac{2}{q}}}{-\dot{H}}\right]^{3/4}=
Q\;\phi^{\alpha_4}, \label{pd}
\end{equation}
where $
Q=\left[\left(\frac{Af}{2}\right)^{8}\left(\frac{C_\phi}{C_\gamma}\right)^{4}\left(\frac{2\,\theta}{C}\right)^{\frac{8(f-1)}{\theta}}\right]^{1/4}$
 and $\alpha_4= \frac{1}{\theta}\left[(3f-4)+\frac{2(1-f)}{q}\right].
$

The scalar spectral index $n_s$ is given by $ n_s -1 =\frac{d
\ln\,{\cal{P}_R}}{d \ln k}$,  where the interval in wave number is
related to the number of e-folds by the relation $d \ln k(\phi)=d
N(\phi)=(H/\dot{\phi})\,d\phi$. From Eqs. (\ref{wr1}) and
(\ref{pd}), we get,
\begin{equation}
n_s=1+I \phi^{\,\alpha_5},\label{nss1}
\end{equation}
where
$
I=\left(\frac{2}{3\,q\,\beta_q^{\frac{2}{q}}}\right)^{1/2}\alpha_4(Af)^{\frac{1}{q}-\frac{3}{2}}
\,(1-f)^{\frac{1}{2}} \,\left(\frac{2\,\theta}{C}\right)^{\frac{1}{\theta}\left[\frac{f-1}{q}+1 -\frac{3f}{2}\right]} $ and
$\alpha_5 =\frac{1}{\theta}\left[\frac{f-1}{q}+1 -\left(\theta +\frac{3f}{2}\right)\right].
$

Note that the scalar spectral index can
be re-expressed in terms of the number of e-folding, $N$. By using
Eqs.(\ref{N1}) and (\ref{al}) we have
\begin{equation}
n_s=1+\frac{3f-4 +\frac{2(1-f)}{q}}{[1+f\,(N-1)]},
\end{equation}
and the value of $f$ in terms of the $n_s$ and $N$ becomes
$$
f=\frac{\frac{2}{q}-(n_s+3)}{N(n_s -1)-(n_s +2)  + \frac{2}{q}}.
$$
In particular, for GB brane world cosmology ( $q=2/3$), $n_s=0.96$ and $N=60$ we obtain that $f\simeq
0.12$.

From Eqs.(\ref{N1}), (\ref{al}), (\ref{pd}) and  (\ref{nss1}), we
can write the parameter $A$ in terms of the particle physics
parameters $C_\gamma$ and $C_\phi$, in the form

\begin{equation}
A=\left(\frac{4\{1+f(N-1)\}^{\frac{\alpha_4}{\alpha_5}}C_\gamma{\cal{P}_R}}{f^{\left(2+
\frac{\alpha_4}{\alpha_5}\right)} C_\phi}\right)^{\alpha_7}\,M
\,,\label{A}
\end{equation}
where
$
M=\left[\frac{3\,q\,\beta_q^{\frac{2}{q}}\,\theta^{2}}{2\,f^{\frac{2}{q}-1\,}\,(1-f)}\right]^{\alpha_7\left[\frac{\alpha_4}
{2\alpha_5}\{\frac{1}{\theta}(\frac{f-1}{q}+1-\frac{3f}{2})-1\}-\frac{f-1}{\theta}\right]}$
, $\alpha_7 = \frac{f}{\alpha_6\left\{(3f-4)+\frac{2(1-f)}{q}\right\}-\alpha_5},
$ and $
\, \alpha_6 = \frac{1}{q}-\frac {1}{2}\left\{3+ \frac{1}{\theta}\left(\frac{2}{q}-1\right)
 \left(\frac{f-1}{q}+1- \frac{3f}{2}\right)\right\} .
$

\begin{figure}[th]
\includegraphics[width=6.0in,angle=0,clip=true]{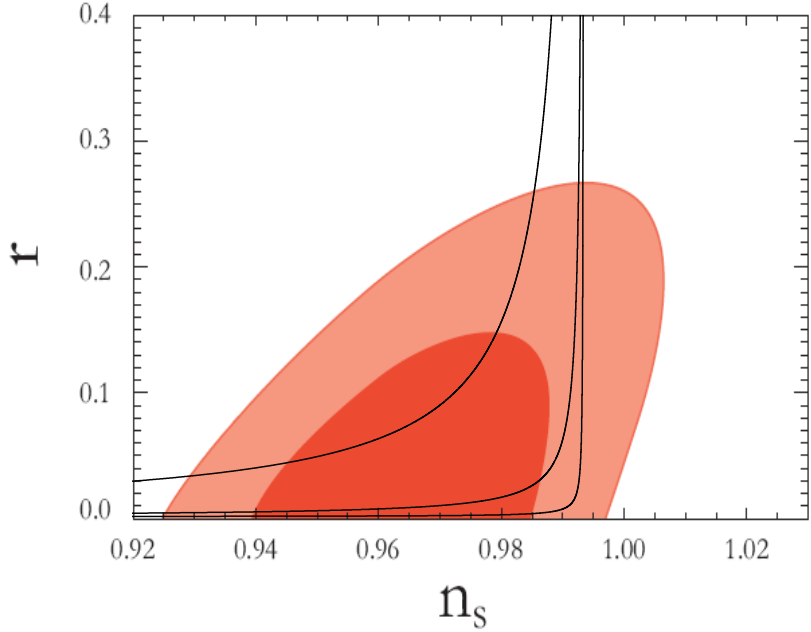}
\caption{Evolution of the tensor-scalar ratio $r$ versus the
scalar spectrum index $n_s$ in the weak dissipative regime, for
the GB cosmology ($q=2/3$) and  three different values of the
parameter $C_\phi$. Here we used $f=3/5$, $\kappa=1$,
$C_\gamma=70$, and $\beta_{2/3}^2=10^{-3} m_p^{-2/3}$.
 \label{rons}}
\end{figure}

In the following we will consider  that the general expression for
the amplitude of scalar perturbations in GB brane world is given
by\cite{pp}
\begin{equation}
\mathcal{P}_{{\mathcal{R}},GB}={\cal{P}_{\cal{R}}}\,\;G_\beta^2(H/\mu)=Q\;\phi^{\alpha_4}\,G_\beta^2(H/\mu),\label{P}
\end{equation}
where  the GB brane world correction is given by
$$
G_\beta^2(x)=\left[\frac{3\,(1+\beta)\,x^2}{2(3-\beta+2\beta\,x^2)\sqrt{1+x^2}+2(\beta-3)}\right]^3,
$$
where $x\equiv \,H\mu$ is a dimensionless measure of energy scale,
and $\beta=4\alpha\mu$. The RS amplification factor is recovered
when $\beta=0$\cite{Maartens:1999hf}.

As it was mentioned in Ref.\cite{Bha} the generation of tensor
perturbations during inflation would  produce  gravitational wave.
In order to confront these models with observations, we need to
consider the $q$-tensor-scalar ratio
$r_q=16\,A_{T,q}^2/A_{S,q}^2$, where the $q$-scalar amplitude is
normalized by $A_{S,q}^2=4\mathcal{P}_{\mathcal{R}}/25$. Here, the
tensor amplitude is given by
\begin{equation}
 A_{T,q}^2=A_{T,
GR}^2\,F_\beta^{2}(H/\mu),\label{At}
\end{equation}
 where  $A_{T,GR}^2$
is the standard amplitude in GR i.e.,
$A_{T,GR}=\sqrt{24}\,\beta_1\,(H/2\pi)$, and the function
$F_\beta$ contains the information about the GB term \cite{pp}
$$
F_\beta^{-2}=\sqrt{1+x^2}-\left(\frac{1-\beta}{1+\beta}\right)
\,x^2\,\sinh^{-1}\left(\frac{1}{x}\right)\;\;\;\;\;(x\equiv\frac{H}{\mu}).
$$

Following, Ref.\cite{Calcagni:2004bh}  we approximate the function
$F_\beta^2\approx F_q^2$, where for the GR regime
$F_{q=1}^2\approx F_\beta^2(H/\mu\ll 1)=1$, for  the RS regime
$F_{q=2}^2\approx F_{\beta=0}^2(H/\mu\gg 1)=3H/(2\mu)$, and
finally for the GB regime $F_{q=2/3}^2\approx F_\beta^2(H/\mu\gg
1)=(1+\beta)H/(2\beta\mu)$. The tensor amplitude up to
leading-order is given by
\begin{equation}
A_{T,q}^{2}=\frac{3q\beta_{q}^{2-2(1-q^{-1})}}{\left(  5\pi\right)  ^{2}}%
\frac{H^{2+2(1-q^{-1})}}{2\zeta_{q}},\label{Ag}
\end{equation}
with $\zeta_{q=1}=\zeta_{q=\frac{2}{3}}=1$\ and \ $\zeta_{q=2}=\frac{2}{3}%
$ \cite{Kim:2004gs}. Finally, the $q$-tensor-scalar ratio from
Eqs.(\ref{P}) and (\ref{Ag}) becomes
\begin{equation}
r_q=16\frac{A_{T,q}^{2}}{A_{S,q}^{2}}==\frac{16}{\zeta_{q}}\frac{(2-f)}{A\,f}\;
\left(\frac{C}{2\,\theta}\right)^{f/\theta}\frac{1}{\phi^{f/\theta}\,\,G_\beta^2(\phi)},\label{r}
\end{equation}
in our cosmological models.





In particular, the Fig.\ref{rons} we show the dependence of the
tensor-scalar ratio $r$ on the spectral index $n_s$ for the GB
regime, where $q=2/3$. From left to right $C_\phi=10^{8}$,
$C_\phi=10^{9}$  and $C_\phi=10^{10}$, respectively.  From
Ref.\cite{Larson:2010gs}, two-dimensional marginalized
 constraints (68$\%$ and 95$\%$ confidence levels) on inflationary parameters
$r$ and $n_s$, the spectral index of fluctuations, defined at
$k_0$ = 0.002 Mpc$^{-1}$. The seven-year WMAP data places stronger
limits on $r$. In order to write down values that relate $n_s$ and
$r$, we used Eqs. (\ref{nss1}) and (\ref{r}).  Also we have used
the values $f=3/5$, $\kappa=1$, $C_\gamma=70$, and
$\beta_{2/3}^2=10^{-5} m_p^{-2/3}$, respectively. From
Eqs.(\ref{N1}) and (\ref{r}), we observed numerically that for the
GB regime and $C_\phi=10^8$, the curve $r=r(n_s)$ (see
Fig.\ref{rons}) for WMAP-7 years enters the 95$\%$ confidence
region for $r\simeq 0.26$, which corresponds to the number of
e-folds, $N\simeq 228$. The curve $r=r(n_s)$  enters the 68$\%$
confidence region  for $r\simeq 0.15$ corresponds to $N\simeq
128$, in this way the GB regime is viable for large values of the
number of e-folds $N$. We also noted that the parameter $C_\phi$,
which is bounded from bellow, $C_\phi>10^{7}$, the GB regime is
well supported by the data as could be seen from Fig.(\ref{rons}).

\section{ The  strong dissipative regime.\label{section3}}

We consider now the case in which $\Gamma$ is large enough for the
system to remain in strong dissipation until the end of inflation,
i.e. $R>1$. From Eqs.(\ref{inf3}) and (\ref{at}), we can obtained
a relation between the scalar field and cosmological times given
by
\begin{equation}
\phi(t)=\phi_0\,\exp[\tau_1\;t^{\,\mu_1}],\label{wr12}
\end{equation}
where $\phi(t=0)=\phi_0$ , $\tau_1$ and $\mu_1$ are defined by
$$
\tau_1= \frac{1}{\mu_1}\left[\frac{2^{7}\,C_\gamma^{3}\,(1-f)}{q
\,
\beta_q^{\frac{2}{q}}}\right]^{1/8}\,\frac{(A\,f)^{\frac{1}{4}(\frac{1}{q}+\frac{3}{2})}}{C_\phi^{1/2}}\,\;\;\;\mbox{and
}
\;\;\;\mu_1=\frac{1}{4}\,\left[\frac{f-2}{2}+(f-1)(\frac{1}{q}+1)\right]+1.
$$
The Hubble parameter as a function of the inflaton field, $\phi$,
result as
\begin{equation}
H(\phi)=A\,f\,\left[\frac{1}{\tau_1}\;\ln(\phi/\phi_0)\right]^{-(1-f)/(\mu_1)}.\label{HH2}
\end{equation}
Without loss of generality we can taken $\phi_0=1$.

From Eq.(\ref{G1}) the dissipation coefficient, $\Gamma$, can be
expressed as a function of the scalar field, $\phi$, as follows
\begin{equation}
\Gamma(\phi)=\frac{\tau_2}{\phi^2}\,\left[\ln(\phi)\right]^{-\mu_2},\label{gg2}
\end{equation}
where $ \tau_2=\left[\frac{(1-f)(Af)^{\frac{2}{q}-1}}{2\,C_\gamma
\, q\, \beta_q^{\frac {2}{q}} }\right]^{3/4} \,
C_\phi\,\,\tau_1^{\mu_2}\;$
\;\;\mbox{and}\;\;\;$\,\mu_2=\frac{3}{4\,
\mu_1}\left[\frac{2(1-f)}{q}+f\right]. $

From Eq.(\ref{pot}) the scalar potential $V(\phi)$, becomes
\begin{equation}
V(\phi)\simeq
\left[\frac{\,A\,f}{\beta_q}\right]^{2/q}\;\left[\frac{1}{\tau_1}\;\ln(\phi)\right]^{-\frac{2}{q}\,(\frac{1-f}{\mu_1})},\label{pot11}
\end{equation}
and as in the previous case,  this kind of potential does not
present a minimum.

In this regime the dimensionless slow-roll parameters  are
\begin{equation}
\epsilon_1^q=-\frac{\dot{H}}{H^2}=\left(\frac{1-f}{A\,f}\right)\;\left[\frac{\tau_1}{\ln(\phi)}\right]^{\frac{f}{\mu_1}},\label{ep1}
\end{equation}
and
\begin{equation}
 \eta_1^q=\left(\frac{2-f}{A\,f}\right)\;\left[\frac{\tau_1}{\ln(\phi)}\right]^{\frac{f}{\mu_1}}.\label{eta2}
\end{equation}

Imposing  the condition $\epsilon_1^q=1$  at the beginning of
inflation (see Ref.\cite{Barrow3}),  the scalar field $\phi$,
takes at this time the value
\begin{equation}
\phi_{1}=\exp\left(\tau_1\;\left[\frac{1-f}{A\,f}\right]^{\frac{\mu_1}{f}}\right).
\label{al22}
\end{equation}

The number of e-folds  becomes given by
\begin{equation}
N=\int_{t_1}^{t_{2}}\,H\,dt=A\,\tau_1 ^{-\frac{f}{\mu_1}}\,
\left[(\ln\phi_{2})^{\frac{f}{\mu_1}}-(\ln\phi_{1})^{\frac{f}{\mu_1}}\right],
\label{N22}
\end{equation}
where Eq.(\ref{wr12}) was used.

In this regime and following  Ref.\cite{Bere2}, we can write
$\delta\phi^2\simeq\,\frac{k_F\,T\,}{2\,\pi^2}$, where the
wave-number $k_F$ is defined by $k_F=\sqrt{\Gamma H}=H\,\sqrt{3
R}> H$, and corresponds to the freeze-out scale at which
dissipation damps out to the thermally excited fluctuations.
From Eqs.(\ref{wr12}) and (\ref{gg2}) we obtained that
\begin{equation}
{\cal{P}_{\cal{R}}}\simeq\frac{1}{2\,\pi^2}\,
\left[\frac{\Gamma^{3}\,H^9}{4\,C_\gamma\,\dot{\phi}^6}\right]^{1/4}\simeq\frac{1}{4\,\pi^2}\,
\left[\frac{\Gamma^{6}\,q^3\,\beta_q^{\frac{6}{q}}\,H^{12-\frac{6}{q}}}{2\,C_\gamma\,(-\dot{H})^3}\right]^{1/4}
\simeq\upsilon_1\,\frac{(\ln\phi)^{\upsilon_3}}{\phi^3}\,,
\label{pd21}
\end{equation}
where
$\upsilon_1=\frac{1}{4\pi^2}\,\left[\frac{q^3\,\beta_q^{\frac{6}{q}}\,\tau_2^6\,(Af)^{9-\frac{6}{q}}\tau_1^{\upsilon_2}}
{2\,C_\gamma\,(1-f)^3}\right]^{1/4},$
$\upsilon_2=24\frac{[2(f+q-1)-3fq]}{[2(2q-1)+f(2+3q)]},$ and   $
\upsilon_3 = 3\,\frac{[f(2+3q)-2(1+2q)]}{[f(2+3q)-2(1-2q)]}.$

From Eq.(\ref{pd21})  the scalar spectral index
$n_s=d\,{\cal{P}_{\cal{R}}}/d\ln k$, is given by
\begin{equation}
n_s\simeq
1-\left[\frac{3\,\ln(\phi)-\upsilon_3}{\gamma_6\,\ln(\phi)^{\gamma_5}}\right],\label{nss2}
\end{equation}
where $\gamma_5 = \frac{8\,f\,q}{2(2q-1)+f(2+3q)},$ $
\gamma_6=\frac{(A\,f)^{\frac{5}{8}-\frac{1}{4q}}}{(1-f)^{1/8}}\,\left[C_\phi\,q\,\beta_q^{\frac{2}{q}}
\left(\frac{1}{2\,C_\gamma\,q\,\beta_q^{\frac{2}{q}}}\right)^{3/4}\right]^{1/2}
\,\tau_1^{\gamma_7}$ and
$\gamma_7=\frac{f(2-5q)+4q-2}{4q-2+f(2+3q)}. $

The scalar spectra index $n_s$ also can be write in terms of the
number of e-folds $N$. Thus, using Eqs. (\ref{al22}) and
(\ref{N22}), we get
\begin{equation}
n_s\simeq 1-f\,A\,\left[\frac{3\,\tau_1
\,[1+f\,(N-1)]^{\frac{1}{\gamma_5}}\,(f\,A)^{-\frac{1}{\gamma_5}}-\upsilon_3
}{\gamma_6\,[1+f(N-1)]\,\tau_1 ^{\gamma_5}} \right].\label{ns2}
\end{equation}

For the strong dissipative regime  we may write the
q-tensor-scalar ratio as
\begin{equation}
r_q=24\,r_0\,\left[\frac{\phi^3}{(\ln
\phi)^{\upsilon_3+2(1-f)/\mu_1}}\right]\,\,\frac{F_\beta^2(x)}{G_\beta^2(x)},
\label{Rkk2}
\end{equation}
where
$r_0=\left[\frac{25\,\beta_1^2\,A^2\,f^2\,\tau_1^{2(1-f)/\mu_1}}{\upsilon_1\,\pi^2}\right]$
and $x=H/\mu$.

\begin{figure}[th]
\includegraphics[width=6.0in,angle=0,clip=true]{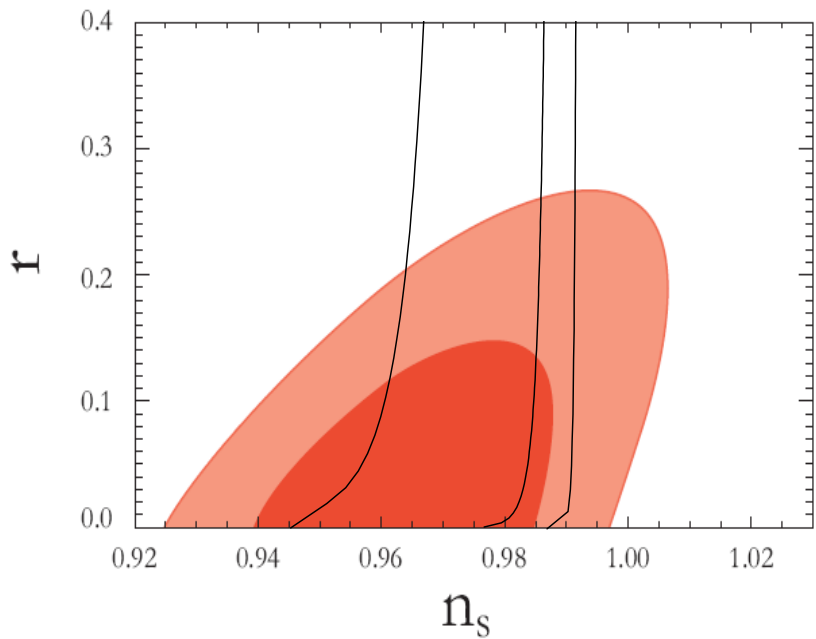}
\caption{Evolution of the tensor-scalar ratio $r$ versus the
scalar spectrum index $n_s$ in the strong dissipative regime, for
three different values of the parameter $C_\phi$ for the GB
regime, where $q=2/3$. Here, we have used $f=3/5$, $\kappa=1$,
$C_\gamma=70$ and  $\beta_{2/3}^2=10^{-5} m_p^{-2/3}$.
 \label{fig2}}
\end{figure}

In particular the Fig.(\ref{fig2})  shows (for the strong
dissipative regime) the dependence of the tensor-scalar ratio on
the spectral index for the GB regime, i.e. $q=2/3$. Here, we have
used different values for the parameter $C_\phi$. From left to
right $C_\phi=10^{6}$, $C_\phi=10^{7}$  and $C_\phi=10^{8}$,
respectively.  In order to write down values that relate $n_s$ and
$r$, we used Eqs. (\ref{nss2}) and (\ref{Rkk2}).  Also we have
used the values $f=3/5$, $\kappa=1$, $C_\gamma=70$, and
$\beta_{2/3}^2=10^{-3} m_p^{-2/3}$, respectively. From
Eqs.(\ref{N22}) and (\ref{Rkk2}), we observed numerically that for
$C_\phi=10^7$, the curve $r=r(n_s)$  for WMAP-7 years enters the
95$\%$ confidence region for $r\simeq 0.26$, which corresponds to
the number of e-folds, $N\simeq 539$. The curve $r=r(n_s)$  enters
the 68$\%$ confidence region  for $r\simeq 0.13$ corresponds to
$N\simeq 467$, in this way the GB regime is viable also for large
values of  $N$, in the strong dissipative regime. We also noted
that the parameter $C_\phi$, which is bounded from bellow,
$C_\phi>10^{6}$, the model is well supported by the data.



\section{Conclusions \label{conclu}}

In this paper we have studied the warm-intermediate inflationary
universe model in braneworld cosmologies, in the weak and  strong
dissipative regimes.  We have also obtained explicit expressions
for the corresponding, dissipation coefficient $\Gamma$,  scalar
potential $V(\phi)$,  the number of e-folds $N$, power spectrum of
the curvature perturbations $P_{\cal R}$, q-tensor-scalar ratio
$r_q$ and scalar spectrum index $n_s$.

In order to bring some explicit results we have taken the
constraint $r_q-n_s$ plane to first-order in the slow roll
approximation. When $\Gamma<3H$ warm inflation occurs in the
so-called weak dissipative regime. In this case, the dissipation
coefficient $\Gamma\propto\phi^{-\alpha_1}$ for intermediate
inflation and the scalar potential $V(\phi)\propto
\phi^{-\alpha_2}$. In particular, we noted that for the GB regime
($q=2/3$) the parameter $C_\phi$, which is bounded from bellow,
$C_\phi>10^7$, the model is well supported by the data as could be
seen from Fig.(\ref{rons}). Here, we have used the WMAP seven year
data, and we have taken the values $f=3/5$, $\kappa=1$,
$C_\gamma=70$, and $\beta_{2/3}^2=10^{-5} m_p^{-2/3}$,
respectively. On the other hand, when $\Gamma>3H$ warm inflation
occurs in the so-called strong dissipative regime. In this regime,
the dissipation coefficient $\Gamma$ present a dependence
proportional to $[\log(\phi)]^{-\mu_2}/\phi^2$ and the scalar
potential $V(\phi)\propto [\ln(\phi)]^{-2(1-f)/(q\mu_1)}$. In
particular, the Fig.(\ref{fig2}) shows that for the values of the
parameter $C_\phi= 10^6$, $10^7$ or $10^8$, the model is well
supported  by the WMAP 7-data, when the values $q=2/3$, $f=3/5$,
$\kappa=1$, $C_\gamma=70$, and $\beta_{2/3}^2=10^{-3} m_p^{-2/3}$,
are taken.

In this paper, we have not addressed the non-Gaussian effects
during warm inflation (see e.g., Refs.\cite{27,fNL}). A possible
calculation from the non-linearity parameter $f_{NL}$, would give
new constrains on the parameters of the model. On the other hand,
when $R<1$ the dissipation does not affect the dynamics of
inflation. When $R>1$ the dissipation does control the dynamics of
the inflaton field. Given that the ratio $R$ will also evolve
during inflation, we may have also models where we start say with
$R<1$ but end in $R>1$, or the other way round. In this paper, we
have not treated these dynamics.  We hope to return to these
points in the near future.

\begin{acknowledgments}
 This work was
supported by COMISION NACIONAL DE CIENCIAS Y TECNOLOGIA through
FONDECYT grant N$^0$ 1090613, and by DI-PUCV  123703.
\end{acknowledgments}



\begin{thebibliography}{99}








\bibitem{warm}
A. Berera,   Phys. Rev. Lett. {\bf 75}, 3218 (1995);  A.~Berera,
  Contemp.\ Phys.\  {\bf 47}, 33 (2006);
M.~Bastero-Gil and A.~Berera,
  Int.\ J.\ Mod.\ Phys.\  A {\bf 24}, 2207 (2009).

\bibitem{agre}I.~G.~Moss,
  Phys.\ Lett.\  B {\bf 154}, 120 (1985);


\bibitem{62526}
L.M.H. Hall, I.G. Moss  and A. Berera,   Phys.Rev.D {\bf 69},
083525 (2004); I.G. Moss,  Phys.Lett.B {\bf 154}, 120 (1985); A.
Berera  and L.Z. Fang, Phys.Rev.Lett. {\bf 74} 1912 (1995).

\bibitem{1126}A. Berera,   Phys. Rev.D {\bf 54},
2519 (1996).

\bibitem{taylorberera}  A. Berera,    Phys. Rev. D {\bf 55},
3346 (1997); J. Mimoso, A. Nunes  and D. Pavon,  Phys.Rev.D {\bf
73}, 023502 (2006); R.~Herrera, S.~del Campo and C.~Campuzano,
  JCAP {\bf 10}, 009 (2006); S. del Campo, R. Herrera  and D. Pavon,
  Phys. Rev. D {\bf 75}, 083518 (2007); S.~del Campo and R.~Herrera,
  Phys.\ Lett.\  B {\bf 653}, 122 (2007); M.~A.~Cid, S.~del Campo and R.~Herrera, JCAP {\bf 10}, 005
  (2006); S.~del Campo and R.~Herrera,
  Phys.\ Lett.\  B {\bf 665}, 100 (2008);   A.~Berera, I.~G.~Moss and R.~O.~Ramos,
  Rept.\ Prog.\ Phys.\  {\bf 72}, 026901 (2009);
 S.~del Campo, R.~Herrera and J.~Saavedra,
  Eur.\ Phys.\ J.\  C {\bf 59}, 913 (2009); R.~Herrera,
  Phys.\ Rev.\  D {\bf 81}, 123511 (2010).



\bibitem{atp} J.~Yokoyama and K.~Maeda,
  Phys.\ Lett.\  B {\bf 207}, 31 (1988).


\bibitem{Barrow1} J. D Barrow,
Phys. Lett. B {\bf 235}, 40 (1990); J. D Barrow and P. Saich,
Phys. Lett. B {\bf 249}, 406 (1990).

\bibitem{Barrow2} J. D Barrow and A. R. Liddle,
Phys. Rev. D {\bf 47}, R5219 (1993).

\bibitem{astro}
  J.~L.~Weiland {\it et al.},
  Astrophys.\ J.\ Suppl.\  {\bf 192}, 19 (2011); C.~L.~Bennett {\it et al.},
  Astrophys.\ J.\ Suppl.\  {\bf 192}, 17 (2011).


  \bibitem{astro2}   N.~Jarosik {\it et al.},
  Astrophys.\ J.\ Suppl.\  {\bf 192}, 14 (2011); E.~Komatsu {\it et al.}  [WMAP Collaboration],
  Astrophys.\ J.\ Suppl.\  {\bf 192}, 18 (2011).





\bibitem{ratio r}
W. H. Kinney, E. W. Kolb, A. Melchiorri and A. Riotto, Phys. Rev.
D {\bf 74}, 023502 (2006); J. Martin and C. Ringeval JCAP {\bf 08}
(2006); F. Finelli, M. Rianna and N. Mandolesi, JCAP {\bf 12} 006
(2006).

\bibitem{Barrow3} J. D. Barrow, A. R. Liddle and C. Pahud, Phys. Rev. D, {\bf 74}, 127305 (2006).

\bibitem{yo4} S.~del Campo and R.~Herrera,
  JCAP {\bf 0904}, 005 (2009).

\bibitem{yo6} S.~del Campo and R.~Herrera,
  Phys.\ Lett.\  B {\bf 670}, 266 (2009).

\bibitem{6} J. Polchinski, String Theory (2 Vols., Cambridge University Press, Cambridge,
1998).

\bibitem{7} J. E. Lidsey, D. Wands, and E. J. Copeland, Phys. Rep. 337, 343 (2000);
M. Gasperini and G. Veneziano, Phys. Rep. 373, 1 (2003).

\bibitem{8}K. Akama, Lect. Notes Phys. 176, 267 (1982); L. Randall and R. Sundrum, Phys. Rev. Lett. 83,
3370 (1999).

\bibitem{BD}D. G. Boulware and S. Deser, Phys.Rev. Lett.
{\bf 55}, 2656 (1985); Phys. Lett. {\bf B 175}, 409 (1986).

\bibitem{KM}T. Kolvisto and D. Mota, Phys. Lett. {\bf B 644},
104 (2007).

\bibitem{ART}I. Antoniadis, J. Rizos and K. Tamvakis, Nucl.Phys.
{\bf B 415}, 497 (1994).

\bibitem{Varios1}S. Mignemi and N. R. Steward, Phys. Rev. D {\bf 47}, 5259 (1993);
 Ch.-M Chen, D. V.
Gal'tsov and D. G. Orlov, Phys. Rev. D {\bf 75}, 084030 (2007).

\bibitem{Varios2}S. Nojiri, S. D. Odintsov and M. Sasaki, Phys. Rev. D {\bf 71},
123509 (2004); G. Gognola, E. Eizalde, S. Nojiri, S. D. Odintsov
and E. Winstanley, Phys. Rev. D {\bf 73}, 084007 (2006).



\bibitem{yo5} R.~Herrera and N.~Videla,
  Eur.\ Phys.\ J.\  C {\bf 67}, 499 (2010).


\bibitem{yo7} S.~del Campo, R.~Herrera, J.~Saavedra and P.~Labrana,
  Annals Phys.\  {\bf 324}, 1823 (2009); S.~del Campo, R.~Herrera and J.~Saavedra,
  Mod.\ Phys.\ Lett.\  A {\bf 23}, 1187 (2008).

\bibitem{Sanyal}A. K. Sanyal, Phys. Lett. {\bf B}, {\bf 645},1
(2007).





\bibitem{Ch} C.Charmousis and J-F. Dufaux, Class.\ Quant.\ Grav.\  {\bf 19}, 4671
(2002).


\bibitem{Davis:2002gn}
  S.~C.~Davis,
  Phys.\ Rev.\  D {\bf 67}, 024030 (2003).


\bibitem{Gravanis:2002wy}
  E.~Gravanis and S.~Willison,
  Phys.\ Lett.\  B {\bf 562}, 118 (2003).



\bibitem{Od}J.~E.~Lidsey, S.~Nojiri and S.~D.~Odintsov,
  JHEP {\bf 0206}, 026 (2002).


\bibitem{Od2}S.~Nojiri, S.~D.~Odintsov and S.~Ogushi,
  Int.\ J.\ Mod.\ Phys.\  A {\bf 17}, 4809 (2002).


\bibitem{Od3}S.~Nojiri, S.~D.~Odintsov and S.~Ogushi,
  Phys.\ Rev.\  D {\bf 65}, 023521 (2002).
\bibitem{pp} J.~F.~Dufaux, J.~E.~Lidsey, R.~Maartens and M.~Sami,
  Phys.\ Rev.\  D {\bf 70}, 083525 (2004).


\bibitem{Calcagni:2004bh}
  G.~Calcagni,
  Phys.\ Rev.\  D {\bf 69}, 103508 (2004).

\bibitem{Tsujikawa:2004dm}
  S.~Tsujikawa, M.~Sami and R.~Maartens,
  Phys.\ Rev.\  D {\bf 70}, 063525 (2004)


\bibitem{Kim:2004hu}
  K.~H.~Kim and Y.~S.~Myung,
  Int.\ J.\ Mod.\ Phys.\  D {\bf 14}, 1813 (2005).


\bibitem{Chung:1999zs}
  D.~J.~H.~Chung and K.~Freese,
  Phys.\ Rev.\  D {\bf 61}, 023511 (2000); C.~J.~Feng and X.~Z.~Li,
  Phys.\ Lett.\  B {\bf 692}, 152 (2010).







\bibitem{26}  I.~G.~Moss and C.~Xiong,
  arXiv:hep-ph/0603266; Y.~Zhang, H.~Li, Y.~Gong and Z.~H.~Zhu,
  arXiv:1103.0718 [astro-ph.CO]; M.~Bastero-Gil, A.~Berera and R.~O.~Ramos,
  arXiv:1008.1929 [hep-ph]; J.~C.~Bueno Sanchez, M.~Bastero-Gil, A.~Berera, K.~Dimopoulos and K.~Kohri,
  JCAP {\bf 1103}, 020 (2011).




\bibitem{28} A.~Berera, M.~Gleiser and R.~O.~Ramos,
  Phys.\ Rev.\  D {\bf 58} 123508 (1998); A.~Berera and R.~O.~Ramos,
  Phys.\ Rev.\  D {\bf 63}, 103509 (2001).




\bibitem{27} J.~C.~Bueno Sanchez, M.~Bastero-Gil, A.~Berera and K.~Dimopoulos,
  Phys.\ Rev.\  D {\bf 77}, 123527 (2008).



\bibitem{Kim:2004gs}
  H.~Kim, K.~H.~Kim, H.~W.~Lee and Y.~S.~Myung,
  Phys.\ Lett.\  B {\bf 608}, 1 (2005).


\bibitem{Liddle} A. Liddle  and D. Lyth, Cosmological inflation and
large-scale structure, 2000, Cambridge University.

\bibitem{B1} A. Berera, Nucl. Phys. B {\bf 585}, 666 (2000).

\bibitem{Maartens:1999hf}
  R.~Maartens, D.~Wands, B.~A.~Bassett and I.~Heard,
  Phys.\ Rev.\  D {\bf 62}, 041301 (2000).

\bibitem{Bha}K. Bhattacharya, S. Mohanty  and A. Nautiyal,
 Phys.Rev.Lett. {\bf 97}, 251301 (2006).

\bibitem{Larson:2010gs}
  D.~Larson {\it et al.},
  Astrophys.\ J.\ Suppl.\  {\bf 192}, 16 (2011)


\bibitem{Bere2} A. Taylor  and A. Berera,   Phys. Rev. D {\bf 69},
083517 (2000).


\bibitem{fNL} S.~Gupta, A.~Berera, A.~F.~Heavens and S.~Matarrese,
  Phys.\ Rev.\  D {\bf 66}, 043510 (2002);  I.~G.~Moss and C.~Xiong,
  JCAP {\bf 0704}, 007 (2007); I.~G.~Moss and T.~Yeomans,
  arXiv:1102.2833 [astro-ph.CO].










\end{thebibliography}
\end{document}